\documentclass[prb,twocolumn,showpacs]{revtex4}
\usepackage{graphicx,epsf}
\begin{document}

\def\tende#1{\,\vtop{\ialign{##\crcr\rightarrowfill\crcr
\noalign{\kern-1pt\nointerlineskip}
\hskip3.pt${\scriptstyle #1}$\hskip3.pt\crcr}}\,}

\title{Magnetoroton instabilities and static susceptibilities in higher Landau levels}
\author{M.~O.~Goerbig and C.~Morais Smith}
\affiliation{D\'epartement de Physique de l'Universit\'e de Fribourg, P\'erolles, 
CH-1700 Fribourg, Switzerland}

\begin{abstract}
We present analytical results concerning the magnetoroton instability in higher Landau levels (LL's) evaluated in the single-mode approximation. The roton gap appears at a finite wave vector, which is approximately independent of the LL index $n$, in agreement with numerical calculations in the composite-fermion picture. However, a large maximum in the static susceptibility indicates a charge-density modulation with wave-vectors $q_0(n)\sim 1/\sqrt{2n+1}$, as expected from Hartree-Fock predictions. We thus obtain a unified description of the leading charge instabilities in all LL's.
\end{abstract}
\maketitle

The complex behavior of two-dimensional electrons in a perpendicular magnetic field remains to be fully understood. If the electrons are restricted to one of the two lowest Landau levels (LL's), the system exhibits the fractional quantum Hall effect (FQHE). This behavior cannot be explained in the framework of the Hartree-Fock approximation (HFA), which predicts a charge-density-wave (CDW) ground state \cite{FP} rather than a homogeneous, incompressible liquid of strongly correlated electrons.\cite{laughlin} On the other hand, HFA does appear to capture the physical properties in higher LL's as shown by recent experiments \cite{exp} at filling factors $\nu=9/2,11/2,...$, which confirm the predictions that the ground state of this system is a unidirectional CDW (``stripes''). \cite{FKS} 

Here we present a method which unifies the physics of electrons in the lowest LL's with that in higher LL's. By studying the low-energy collective excitations in the single-mode approximation (SMA), we show that the magnetoroton minimum, which indicates an instability of the incompressible liquid towards Wigner crystallization in the lowest LL's, \cite{GMP,MG} in fact indicates an incipient CDW instability in higher LL's. In order to observe this feature, it is not sufficient to evaluate the energy dispersion of the collective excitations, as performed in previous works, \cite{MG,scarola} but one must consider the static susceptibility. As we will show, the dispersion always presents a magnetoroton minimum at a wave vector $q\sim 1$ (in units of the inverse magnetic length $l_B$), essentially independent of the LL index $n$. However, the static susceptibility exhibits a maximum at $q\sim 1$ for the lowest LL's, a signature of the Wigner crystal (WC) instability, whereas the maxima in higher LL's arise at $q_0(n)\sim 1/\sqrt{2n+1}$, a feature predicted in the HFA for the CDW phase. \cite{FKS}

The FQHE phenomenon at filling factors $\nu=1/(2s+1)$, with $s$ being an integer, was first explained by Laughlin \cite{laughlin} using trial wave functions. These many-body wave functions describe electrons attached to $2s+1$ vortices, and this attachment leads to a minimization of the Coulomb energy, which is the only energy scale in the problem. Later, Jain \cite{jain} proposed a generalization of the trial wave functions, which describe the FQHE at filling factors $p/(2ps\pm 1)$, where the most prominent plateaus are observed experimentally ($p$ is an integer). The FQHE is understood as an integer quantum Hall effect of composite fermions (CF's) subject to a reduced magnetic field and filling $p$ CF-LL's. These new particles are interpreted as electrons bound to $2s$ vortices, and the Laughlin states are obtained when the CF's populate the lowest CF-LL.

If one neglects LL mixing, the underlying Hamiltonian describing the low-energy physics of spin-polarized electrons in any LL $n$ is given in terms only of the Coulomb interaction by
\begin{equation}
\label{equ002}
\hat{H}_n=\sum_{\vec{q}}v(q)\left[F_n(q)\right]^2\bar{\rho}(-\vec{q})\bar{\rho}(\vec{q}),
\end{equation}
where $v(q)=2\pi e^2/(\epsilon q)$ is the Fourier transformation of the Coulomb interaction and $F_n(q)=L_n(q^2/2)\exp(-q^2/4)$ is the form factor in terms of Laguerre polynomials arising from the electron wave functions in the $n$th LL. The operators $\bar{\rho}(\vec{q})$ are the projected Fourier components of the electron density, and obey the commutation relations
$$
[\bar{\rho}(\vec{q}),\bar{\rho}(\vec{k})]=2i\sin\left(\frac{(\vec{q}\times\vec{k})_z}{2}\right)\bar{\rho}(\vec{q}+\vec{k}).
$$
LL mixing in the limit where the cyclotron gap $\hbar eB/m$ is large compared to the characteristic Coulomb energy $e^2/(\epsilon l_B)$ may be taken into account in the form of a screened interaction potential $v(q)=2\pi e^2/\epsilon(q) q$ with the dielectric function $\epsilon(q)$. \cite{AG} This modification is significant only in the vicinity of a characteristic wave vector $q=2.4/\sqrt{2n+1}$. In the following we will neglect this screening because it is unimportant in determining the behavior of electrons in higher LL's. \cite{fogler} Note that the Hamiltonian (\ref{equ002}) allows one to describe electrons in any LL $n$ in terms of electrons in the lowest LL interacting via a modified potential $v_n(q)=v(q)[F_n(q)]^2$.

To describe the collective excitations of electrons in the FQHE regime, Girvin {\sl et al.} \cite{GMP} used the SMA, in which the excited state is given by
$$
|\psi_{\vec{q}}^{2s+1}\rangle=\bar{\rho}(\vec{q})|\Omega_{2s+1}\rangle. 
$$
The energy dispersion of the excitations is
$$
\Delta_{2s+1}(q)=\frac{\langle \psi_{\vec{q}}^{2s+1}|\hat{H}-E_0^{2s+1}|\psi_{\vec{q}}^{2s+1}\rangle}{\langle \psi_{\vec{q}}^{2s+1}|\psi_{\vec{q}}^{2s+1}\rangle},
$$
where $E_0^{2s+1}$ is the energy of the Laughlin state $|\Omega_{2s+1}\rangle$. This energy remains finite at any wave vector, as expected for an incompressible ground state, but has a minimum at a nonzero value of the wave vector. This magnetoroton minimum is interpreted as the precursor of a WC instability because it matches the reciprocal-lattice vector of the WC. The WC is expected to be the true ground state at very low filling factors $\nu$, where the average distance between electrons exceeds the spatial extent $l\sim l_B$ of the single-particle wave functions. With increasing $s$, the gap at the magnetoroton minimum decreases, indicating that the WC becomes energetically more favorable. But in the SMA the gap never vanishes, so one would expect the Laughlin state to be the true ground state even at very low filling factors (large $s$). However, Kamilla and Jain \cite{KJ} have shown that the SMA overestimates the gap at the magnetoroton minimum and fails at larger wave vectors. They compare the SMA to a description of the collective excitations as magnetoexcitons of CF's. The excited state is given by \cite{KWJ}
$$
|\psi_{\vec{q}}^{\rm CF-exc}\rangle={\mathcal{P}}\rho_{\vec{q}}^{p\rightarrow p+1}|\Omega_{p,s}\rangle,
$$
where $|\Omega_{p,s}\rangle$ is the CF ground state, in which CF's containing $2s$ vortices populate $p$ CF-LL's, and $\rho_{\vec{q}}^{p\rightarrow p+1}$ creates a particle-hole pair of wave vector $\vec{q}$, with the particle situated in the ($p+1$)th CF-LL and the hole in the $p$th. $\mathcal{P}$ is the projection operator to the lowest LL. Even if it is not evident how to relate the phononlike excitations introduced in the SMA to the CF magneto-excitons, the two have been shown to coincide in the limit $q\rightarrow 0$. \cite{KWJ} In the CF picture one finds that for $\nu=1/9$ and lower filling factors the collective excitations at the magnetoroton minimum have a lower energy than the Laughlin state. This indicates that the WC is the true ground state of the system in the dilute limit. \cite{KJ}

The question of whether the FQHE states in higher LL's become unstable towards a CDW was discussed by Scarola {\sl et al.} \cite{scarola} using similar arguments as for the WC instability. The authors found that the CF's with two vortices ($^2$CF's) are unstable for $n\geq2$, because the CF excitons have a lower energy. However, the instability is still found at wave vectors of order 1, and not at the characteristic wave-vectors $q_0(n)\sim 1/\sqrt{2n+1}$ as expected for the CDW. Their second remarkable result was that excitons of CF's with 4 or 6 vortices ($^4$CF's or $^6$CF's) show a finite gap at all wave vectors for the LL's $n=2$ and $n=3$. This supports the conjecture that the FQHE could be found in ultrahigh mobility samples around $\bar{\nu}=1/4$ or $\bar{\nu}=1/6$ in higher LL's, where $\bar{\nu}$ now denotes the filling factor of the last LL. \cite{morf}

\begin{figure}
\epsfysize+6cm
\epsffile{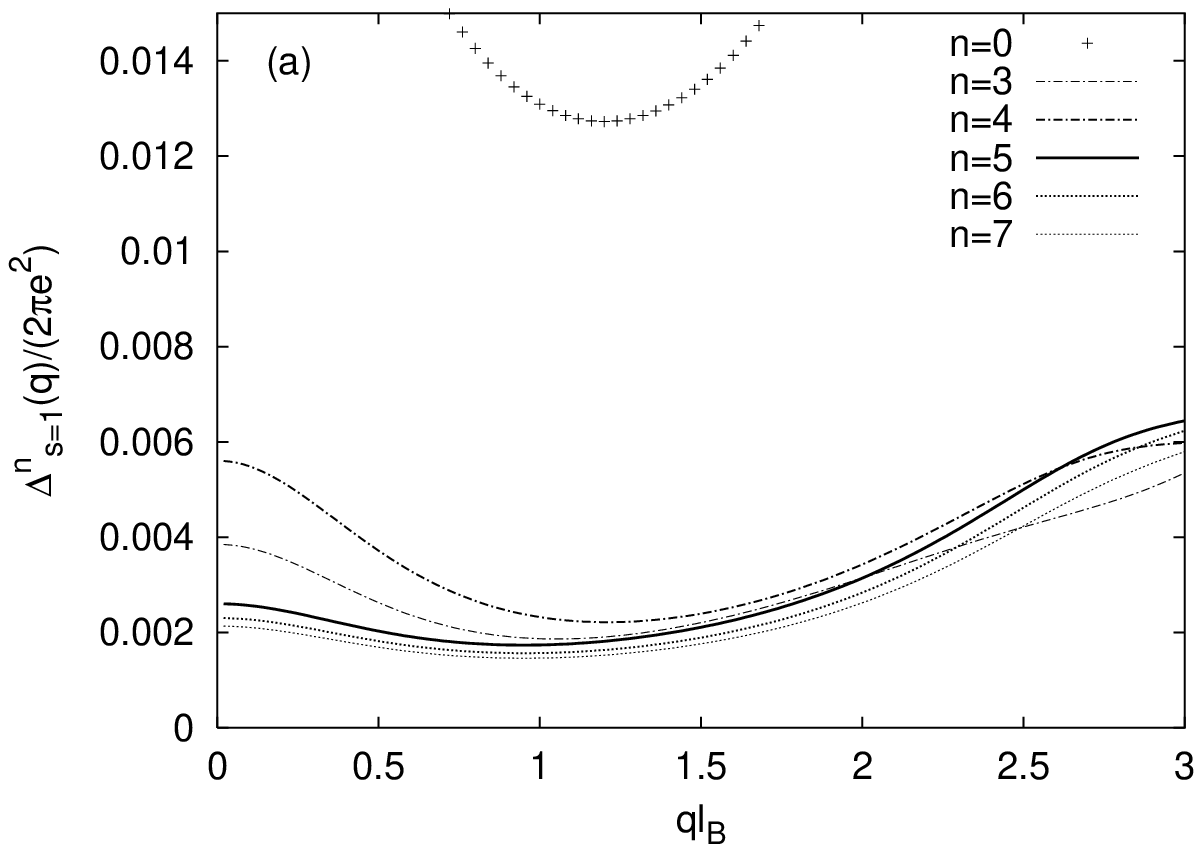}
\epsfysize+6cm
\epsffile{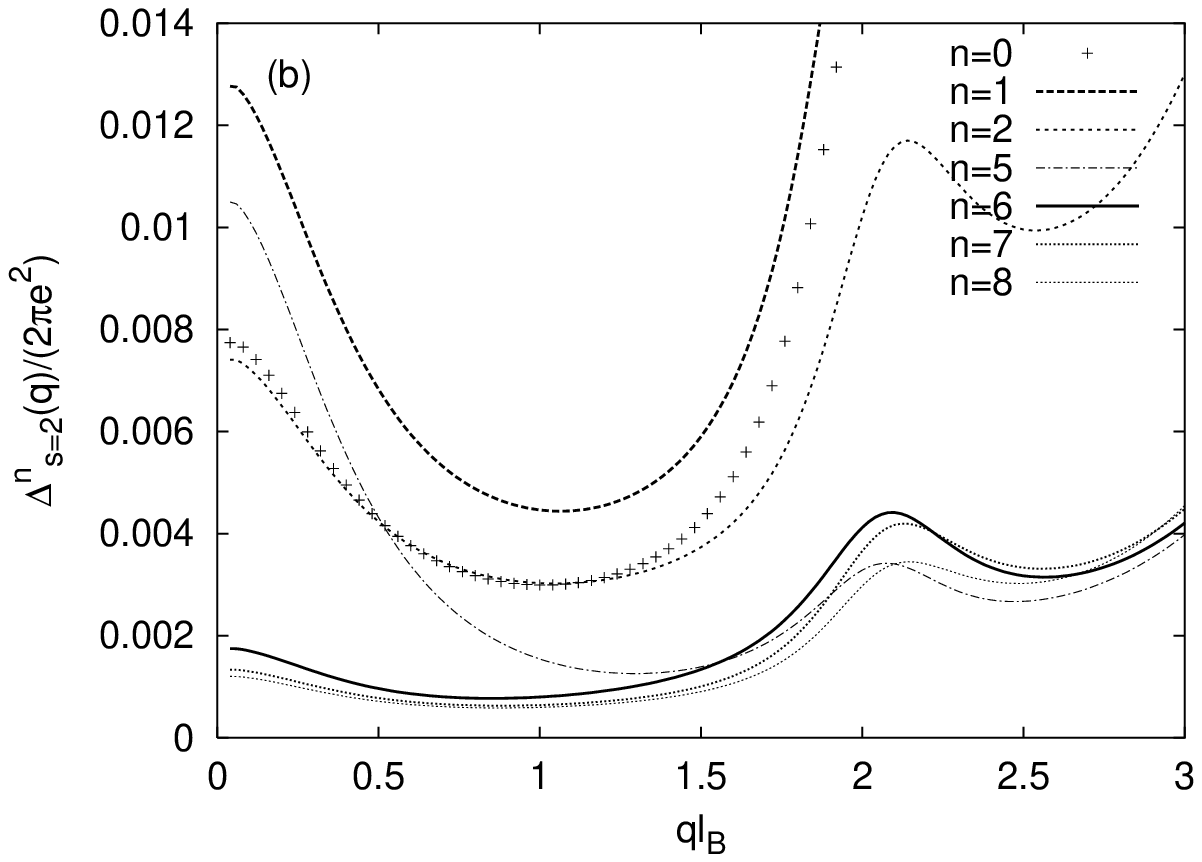}
\caption{Energy dispersions for $\bar{\nu}=1/3$ (a) and $\bar{\nu}=1/5$ (b).}
\label{fig1}
\end{figure}

Here, we present analytical results for the SMA in higher LL's. Studies for $n=1$ were first performed numerically by MacDonald and Girvin. \cite{MG} The phononlike collective excitations in a higher LL are described by the wave function
$$
|\psi_{n,\vec{q}}^{2s+1}\rangle=F_n(q)\bar{\rho}(\vec{q})|\Omega_{2s+1}\rangle, 
$$
where we have made use of the projection scheme presented above. One finds for the energy dispersion
\begin{eqnarray}
\label{equ008}
\nonumber
\Delta_{2s+1}^{(n)}(q)&=&2\sum_{\vec{k}}v(k)\left[F_n(k)\right]^2\sin^2\left(\frac{(\vec{q}\times\vec{k})_z}{2}\right)\\
\nonumber
&&\times\frac{e^{|\vec{q}+\vec{k}|^2/2}\overline{s_{2s+1}(\vec{q}+\vec{k})}-e^{k^2/2}\overline{s_{2s+1}(\vec{k})}}{e^{q^2/2}\overline{s_{2s+1}(\vec{q})}},
\end{eqnarray}
where $\overline{s_{2s+1}(\vec{q})}$ is the projected static structure factor of the ($2s+1$)-Laughlin state. The structure factor is the Fourier transformation of the pair distribution function, which may be calculated using Monte Carlo integration. \cite{GMP} The latter may be expressed as the series \cite{girvin}
\begin{eqnarray}
\label{equ009}
\nonumber
g_{2s+1}(r)&=&1-e^{-r^2/2}\\
\nonumber
&&+\sum^{\infty}_{m=0}\frac{2}{(2m+1)!}c_{2m+1}^{2s+1}\left(\frac{r^2}{4}\right)^{2m+1} e^{-r^2/4},
\end{eqnarray}
which leads to the formula
\begin{eqnarray}
\label{equ010}
\nonumber
\overline{s_{2s+1}(q)}&=&\frac{2s}{2s+1}e^{-q^2/2}+\\
\nonumber
&&\frac{4}{2s+1}\sum_{m=0}^{\infty}c_{2m+1}^{2s+1}L_{2m+1}(q^2)e^{-q^2}
\end{eqnarray}
for the projected static structure factor. In contrast to Ref.~5, where the expansion coefficients $c_{2m+1}^{2s+1}$ were obtained by a fit to the Monte Carlo results, here we use the following sum rules as determining equations for the coefficients,
\begin{eqnarray}
\label{equ011}
\nonumber
&&\sum^{\infty}_{m=0}c_{2m+1}^{2s+1}=-\frac{s}{2},\\
\nonumber
&&\sum^{\infty}_{m=0}(2m+2)c_{2m+1}^{2s+1}=-\frac{s}{4},\\
\nonumber
&&\sum^{\infty}_{m=0}(2m+3)(2m+2)c_{2m+1}^{2s+1}=\frac{s^2}{2},\\
\nonumber
&&c_{2m+1}^{2s+1}=-1\qquad{\rm for}~m<s.
\end{eqnarray}
Coefficients with an index $m\geq M=3+s$, which is the maximal index one can determine in this procedure, are set to zero. The first three sum rules are due to charge neutrality, perfect screening, and a compressibility sum rule, respectively. The last condition is given by the repulsion of the electrons at short distances, where $g(r)\sim r^{2(2s+1)}$. \cite{girvin} The results obtained in the lowest LL for pair distribution function, projected structure factor, and energy dispersion (see curves in crosses in Fig.~\ref{fig1}) are in good agreement with Monte Carlo calculations.

In higher LL's (see Fig.~\ref{fig1}), we obtain the following results: for $\bar{\nu}=1/3$ ($s=1$), the gap at the roton minimum is largest in the lowest LL ($n=0$), $n=1$ and $n=2$ determine an intermediate region (not shown) where the SMA breaks down, and for $n\geq 3$, the roton gap decreases continuously. At $\bar{\nu}=1/5$ ($s=2$), the roton gap {\sl increases} from $n=0$ to $n=1$, the intermediate region is found at $n=3$ and $n=4$, and for $n\geq 5$, we observe the decrease of the gap. The same qualitative behavior is found for $\bar{\nu}=1/7$ ($s=3$) with a maximal roton gap in the $n=3$ LL, the breakdown of the SMA at $n=4$ and $n=5$, and the final decrease for $n\geq6$ (figure not shown). For $\bar{\nu}=1/5$ and $\bar{\nu}=1/7$, the roton gap first increases with increasing $n$ [see the $n=0$ and $n=1$ curves in Fig.~\ref{fig1}(b)] due to the fact that the system is further away from the WC regime. The WC transition appears at  $\bar{\nu}_{WC}(n)\sim 1/(2n+1)$ because the average distance between electrons must be considerably larger than the size of the wave functions, which scales as $\sqrt{2n+1}$ in higher LL's. Therefore, the WC instability indicated by the roton gap is less pronounced.

\begin{figure}
\epsfysize+6cm
\epsffile{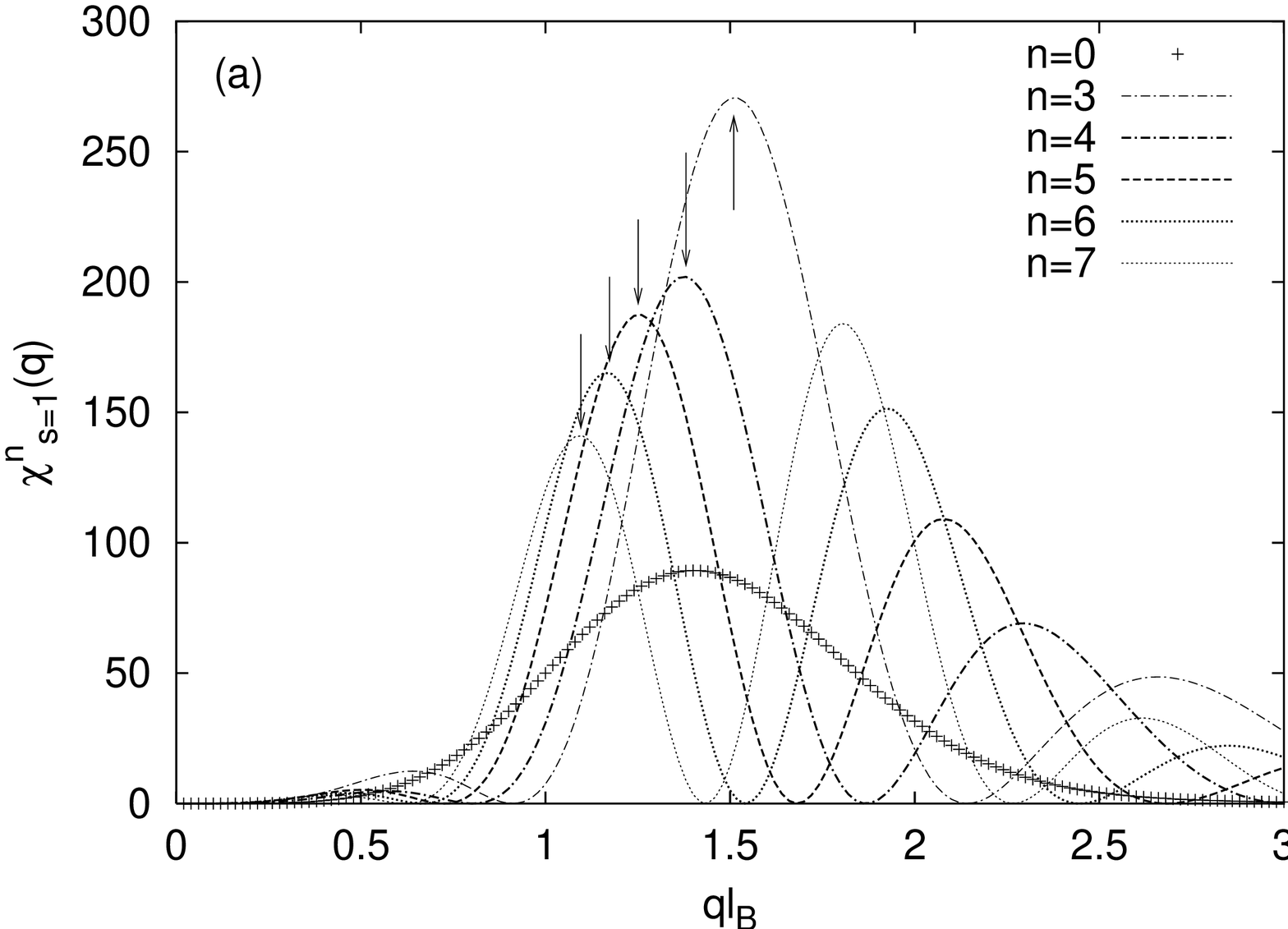}
\epsfysize+6cm
\epsffile{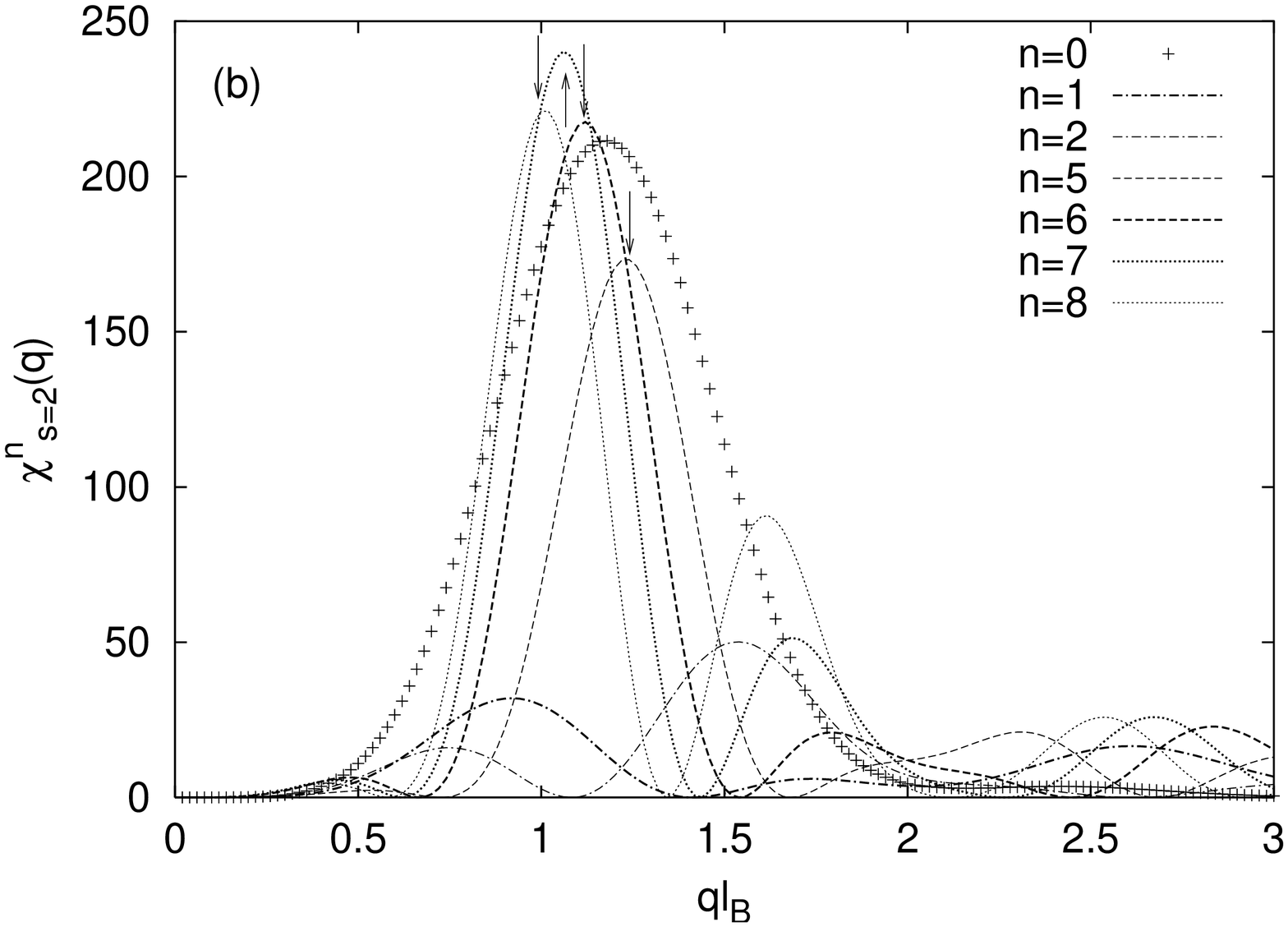}
\epsfysize+6cm
\epsffile{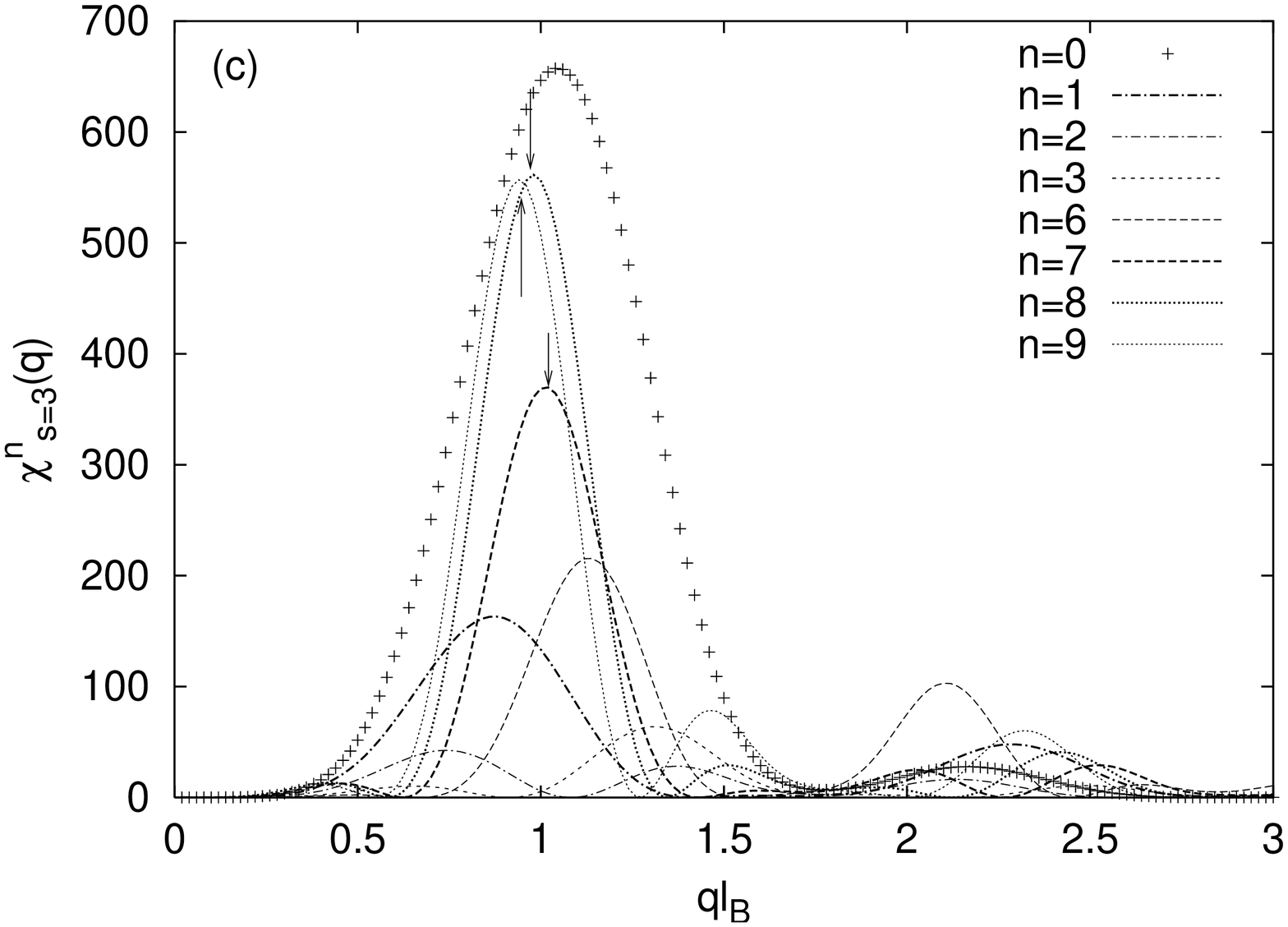}
\caption{Static susceptibilities for $\bar{\nu}=1/3$ (a), $\bar{\nu}=1/5$ (b), and $\bar{\nu}=1/7$ (c).}
\label{fig2}
\end{figure}

For $\bar{\nu}=1/5$, we find a second minimum in the dispersion relations at $q\sim 2.5$, see Fig. 1(b). This minimum is already present in the lowest LL. \cite{MacDo} However, the SMA is known to be less reliable at larger wave vectors, and for this reason no results were shown in Ref.~5 at $q>2.0$ for $\bar{\nu}=1/5$. The second minimum also arises in numerical calculations within the CF picture, which is more accurate in the large-$q$ limit. \cite{KJ}

Note that we do not find the signature of a CDW state in the dispersion relation of the collective excitations, because the magnetoroton minimum occurs at wave-vectors $q\sim 1$, independent of the LL. This is in agreement with the results found in the CF picture. \cite{scarola} The fact that we obtain the most stable $\bar{\nu}=1/7$ FQHE state in the $n=3$ LL, and not for $n=2$, \cite{scarola} is probably due to the proximity to the intermediate region in which the SMA becomes less reliable. Breakdown of the SMA in the intermediate region is expected to be due to a numerical effect, the origin of which is as follows: the gap at $q=0$
\begin{equation}
\label{equ012}
\Delta_{2s+1}^{(n)}(q=0)\propto \sum^{\infty}_{m=0}c_{2m+1}^{2s+1} I_{2m+1}^n
\end{equation}
may be calculated analytically using the integrals 
\begin{eqnarray}
\label{equ013}
\nonumber
I_k^n&=&\int_0^{\infty}dq\left\{
k(k-1)L_{k-2}(q^2)+k(q^2-2k)L_{k-1}(q^2)\right.\\
\nonumber
&+&\left.\left[\frac{q^4}{4}-(k+1)q^2+k(k+1)\right]L_k(q^2)\right\} \left[F_n(q)\right]^2 e^{-\frac{q^2}{2}}.
\end{eqnarray}
Inspection of $I_{2m+1}^n$ as a function of $m$ reveals that the integrals are initially negative, and decrease until a minimum is reached around $n\sim m$. $I_{2m+1}^n$ then jumps to large positive values before decreasing again towards zero. It is therefore in the region $n\sim m$ that the gap at $q=0$ [Eq. (\ref{equ012})] is most sensitive to approximations made in determining the coefficients $c_{2m+1}^{2s+1}$, either numerically or in the present approach. Because we have set $c_{2m+1}^{2s+1}=0$ for $m>M$, it is in the intermediate region defined by $n\sim M$ that the results of the SMA become unreliable even in the small-$q$ limit.

However, the essential aspect of this work is that even if one does not find any signature of the CDW in analyzing the position of the magnetoroton minimum, it can in fact be seen in the static susceptibility calculated in the SMA,
$$
\chi_{2s+1}^n(q)=-2\frac{s^n_{2s+1}(q)}{\Delta^n(q)}=-2\frac{\overline{s_{2s+1}(q)}[L_n(q^2/2)]^2}{\Delta^n(q)},
$$
where $s^n_{2s+1}(q)$ is the static structure factor projected to the LL $n$. Girvin {\sl et al.} \cite{GMP} have already emphasized that the maximum arising in the static susceptibility also indicates the WC instability in the lowest LL. Because in higher LL's the expression for $\chi_{2s+1}^n(q)$ is essentially the same as for the lowest LL multiplied by the Laguerre polynomials of the form factor, the position of the maximum now scales as $q_0(n)\sim 1/\sqrt{2n+1}$ as expected for a CDW instability (see arrows in Fig.~\ref{fig2}). 

Note that this interpretation of the CDW in higher LL's as arising from the overlap between wave functions described by the form factor $F_n(q)$ is in agreement with the ground-state calculations in the HFA. \cite{FKS} At the wave vectors where the Laguerre polynomials have nodes, the repulsive Hartree potential vanishes and only the attractive exchange interaction remains finite. This gives rise to an effective attraction between electrons in the topmost LL at the wave-vector $q_0(n)$ of the CDW.

In conclusion, we have found a tendency of the FQHE states described by Laughlin's wave functions to form a CDW in higher LL's. The analytical calculations were performed within the SMA. The tendency toward CDW formation cannot be observed if one investigates the magnetoroton gap in the dispersion of phononlike collective excitations, but becomes apparent only in the static susceptibility, which exhibits maxima at the expected CDW wave-vectors $q_0(n)\sim1/\sqrt{2n+1}$. Our results thus confirm the prediction of CDW formation in higher LL's in the framework of \cite{FKS} HFA and provide a unified description for the WC and CDW instabilities of all LL's.

We would like to acknowledge fruitful discussions with D.~Baeriswyl, A.~Caldeira, J.~Jain, P.~Lederer, A.~MacDonald, R.~Morf, D.~Pfannkuche, and M.~Rosenau~da~Costa. We further thank B.~Normand for a careful reading of the manuscript. This work was supported by the Swiss National Foundation for Scientific Research under grant No.~620-62868.00.


\begin{references}

\bibitem{FP}H.~Fukuyama, P.~M.~Platzman, and P.~W.~Anderson, Phys.~Rev.~B {\bf 19}, 5211 (1979).

\bibitem{laughlin}R.~B.~Laughlin, Phys.~Rev.~Lett. {\bf 50}, 1395 (1983).

\bibitem{exp} M.~P.~Lilly, K.~B.~Cooper, J.~P.~Eisenstein, L.~N.~Pfeiffer, and K.~W.~West, Phys.~Rev.~Lett. {\bf 82}, 394 (1999); R.~R.~Du, D.~C.~Tsui, H.~L.~Stormer, L.~N.~Pfeiffer, K.~W.~Baldwin, and K.~W.~West, Solid~State~Commun. {\bf 109}, 389 (1999).

\bibitem{FKS}A.~A.~Koulakov, M.~M.~Fogler, and B.~I.~Shklovskii, Phys.~Rev.~Lett. {\bf 76}, 499 (1996);
M.~M.~Fogler, A.~A.~Koulakov, and B.~I.~Shklovskii, Phys.~Rev.~B {\bf 54}, 1853 (1996);
R.~Moessner and J.~T.~Chalker, {\sl ibid.} {\bf 54}, 5006 (1996).

\bibitem{GMP}S.~M.~Girvin, A.~H.~MacDonald, and P.~M.~Platzman, Phys.~Rev.~B {\bf 33}, 2481 (1986).

\bibitem{MG}A.~H.~MacDonald and S.~M.~Girvin, Phys.~Rev.~B {\bf 33}, 4009 (1986).

\bibitem{scarola}V.~W.~Scarola, K.~Park, and J.~K.~Jain, Phys.~Rev.~B {\bf 62}, R16\,259 (2000).

\bibitem{jain}J.~K.~Jain, Phys.~Rev.~Lett. {\bf 63}, 199 (1989); Phys.~Rev.~B {\bf 41}, 7653 (1990).

\bibitem{AG}I.~L.~Aleiner and L.~I.~Glazman, Phys.~Rev.~B {\bf 52}, 11\,296 (1995).

\bibitem{fogler}M.~M.~Fogler and  A.~A.~Koulakov, Phys.~Rev.~B {\bf 55}, 9326 (1997).

\bibitem{KJ} R.~K.~Kamilla and J.~K.~Jain,  Phys.~Rev.~B {\bf 55}, R13\,417 (1997).

\bibitem{KWJ} R.~K.~Kamilla, X.~G.~Wu, and J.~K.~Jain, Phys.~Rev.~B {\bf 54}, 4873 (1996).

\bibitem{morf}R.~Morf and N.~d'Ambrumenil, Phys.~Rev.~Lett. {\bf 74}, 5116 (1995); R.~Morf (private communication).

\bibitem{girvin}S.~M.~Girvin, Phys.~Rev.~B {\bf 30}, 558 (1984).

\bibitem{MacDo}A.~H.~MacDonald (private communication).

\end{references}
\end{document}